\documentstyle[12pt]{article}
\newcommand{\Sidec}{\mbox{$\Sigma ^+ \rightarrow p \gamma $}}

\begin{document}
\title{The puzzle of the quark model: Weak radiative hyperon decays
\footnote{
To appear in the Proceedings of the conference "Production and Decay
of Hyperons, Charm and Beauty Hadrons", Strasbourg, France, September
5-8, 1995
}}
\author{
Report No. 1702/PH\\
\\
\\
\\
P. \.{Z}enczykowski\\
\\
Dept.
of Theor. Physics \\
Institute of Nuclear Physics\\
Radzikowskiego 152,
31-342 Krak\'ow, Poland\\
\\ }
\maketitle
\begin{abstract}
Weak radiative hyperon decays present us with a long-standing puzzle,
namely
the question of validity of a hadron-level theorem proved by Hara.
We briefly discuss the conflict between expectations based on Hara's
theorem and experiment as well as the way in which the quark model
evades the theorem.  Violation of Hara's theorem in the quark model
is traced back to the issue of hadron compositeness and
the nonequivalence
of standard ways of imposing gauge-invariance condition at quark and hadron
levels. This suggests that our understanding of nonlocal composite nature
of hadrons may require some important change.
\end{abstract}
\newpage
\section{INTRODUCTION}
Weak radiative hyperon decays (WRHD's) have proved to be a challenge
to both theorists and experimenters.  Experimental difficulties
result from small branching ratios ($\approx 10^{-3}$) of WRHD's and
their copious photon backgrounds.
Long history of unsuccessful theoretical approaches
has led theorists to view the problem of WRHD's as "a long-standing
discrepancy" \cite{ref:Gaillard}, an "unsolved puzzle"
\cite{ref:Balitsky} or "the long-standing $\Sigma ^+ \rightarrow p \gamma $
puzzle"
\cite{ref:Singer}.
Recently, their actual status
has been extensively reviewed by J. Lach and the
author \cite{ref:Lach}.
It is presented here in brief.

\section{THE CONFLICT}
WRHD's are rare strangeness-changing decays of
hyperons into other ground-state baryons plus a photon.
There are five experimentally observed WRD of ground-state octet
baryons:
$\Sigma ^+ \rightarrow p \gamma $, $\Lambda
\rightarrow n
\gamma
$,
$\Xi ^0 \rightarrow \Sigma ^0 \gamma $,
$\Xi ^0 \rightarrow \Lambda \gamma $, $\Xi ^- \rightarrow
\Sigma
^- \gamma $.

Theoretical problems manifest themselves most clearly in the description
of the \Sidec\ decay.  This particular decay should satisfy a fairly
fundamental theorem proved by Hara \cite{ref:Hara}.
It is therefore extremely interesting that
\begin{itemize}
\item there exists a conflict between experiment and expectations based
on Hara's theorem
\item the quark model evades Hara's theorem in a strange and
thought-provoking way.
\end{itemize}
Hara's theorem, proved at hadron level, reads:
\begin{quote}
{\em Parity-violating amplitude $A$ of the \Sidec\ decay
vanishes in exact $SU(3)$-flavour symmetry.}
\end{quote}
For a nonzero parity-conserving amplitude $B$ one then expects
decay asymmetry
\begin{equation}
\alpha = \frac{2 Re(A^* B)}{|A|^2 + |B|^2}
\end{equation}
to be small
since $SU(3)$ is usually broken weakly.

Current experimental evidence, summarized in Fig.1, shows beyond
any doubt that asymmetry in question is {\em large} (and
{\em negative}).
The most recent number, coming from the E761 experiment performed
at Fermilab \cite{ref:Foucher}, is based on nearly 35 thousand events.
\\
\setlength {\unitlength}{1.2pt}
\begin{picture}(250,135)
\put(80,0){\begin{picture}(150,135)
\put(8,28){-1.5}
\put(8,53){-1.0}
\put(8,78){-0.5}
\put(8,103){\phantom{-}0.0}

\put(0,65){$\alpha $}

\put(18,18){1965}
\put(58,18){1975}
\put(98,18){1985}
\put(138,18){1995}

\put(80,5){Year}

\put(25,30){\begin{picture}(120,100)

\multiput(0,75)(20,0){6}
{\line(1,0){20}
 \line(0,-1){2}}

\multiput(0,0)(20,0){6}
{\line(1,0){20}
\line(0,1){2}}

\multiput(0,0)(0,5){15}
{\line(0,1){5}
 \put(0,0){\line(1,0){2}}}

\multiput(120,0)(0,5){15}
{\line(0,1){5}
\put(0,0){\line(-1,0){2}}}

\multiput(0,37)(6,0){20}{\line(1,0){3}}

\put(16,23.5){\circle*{3}}
\put(16,2.5){\line(0,1){47}}
\put(60,48.5){\circle*{3}}
\put(60,30.5){\line(0,1){37}}
\put(88,32){\circle*{3}}
\put(88,25.5){\line(0,1){13}}
\put(108,39){\circle*{3}}
\put(108,34.5){\line(0,1){9}}

\end{picture}}
\end{picture}}
\end{picture}
\\
\begin{center}
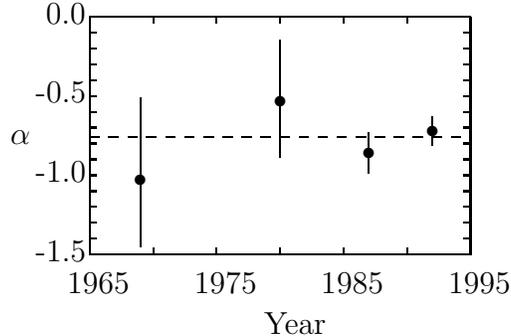
Fig. 1.  History of measurement of $\Sigma ^+ \rightarrow p \gamma $
asymmetry parameter.
\end{center}

A standard first reaction
to the above disagreement between experiment and theoretical expectations
is to say that in this case
$SU(3)$-breaking is perhaps stronger than elsewhere.
In reality the
situation is much more involved and delicate: in 1983 Kamal and
Riazuddin showed \cite{ref:KR} that Hara's theorem
is {\em violated in the quark model} {\em also in the SU(3) limit}.
Explanation of this astonishing result was proposed in
1989 by the author \cite{ref:Zen89}.

Since the quark model violates Hara's theorem even in the $SU(3)$
limit, our attention must be focussed on other assumptions needed
in its proof.  However, the only apparent other
assumptions are:
\\
\phantom{xxxxxxx}1. gauge invariance,
\\
\phantom{xxxxxxx}2. CP-invariance.

Gauge invariance requires that in the most general hadron-photon
parity violating coupling
\begin{equation}
\overline{\Psi }_p \gamma _5
(\gamma _{\mu }F_1(q^2)+q_{\mu }F_2(q^2)+F_3(q^2)\sigma _{\mu \nu }
q^{\nu})
\Psi _{\Sigma ^+ }A^{\mu }
\end{equation}
one has $F_1(0)=0$ and, consequently,
for real, transverse, final photons ($q^2=q_{\mu }A^{\mu }=0$)
only the $F_3$ term contributes.

CP-invariance (which relates $p \leftrightarrow \overline{p}$,
$\Sigma ^+ \leftrightarrow \overline{\Sigma }^-$)
requires that full coupling of the $p, \Sigma ^+ $ initial baryons
and the $\Sigma ^+, p$ final baryons to real photons is
\begin{equation}
F_3(q^2)(\overline{\Psi }_p \gamma _5 \sigma _{\mu \nu }\Psi _{\Sigma ^+}
- \overline{\Psi }_{\Sigma ^+ } \gamma _5 \sigma _{\mu \nu }\Psi _p)
q^{\nu } A^{\mu }
\end{equation}
which is {\em antisymmetric} under $\Sigma ^+ \leftrightarrow p$
interchange.  Since the weak Hamiltonian is {\em symmetric}
under $s \leftrightarrow d$ ($\Sigma ^+ \leftrightarrow p$) interchange
($SU(3)$ limit) we must have $F_3 = 0$ and, consequently, the
parity-violating
\Sidec\ amplitude vanishes.

One might therefore expect that the quark-model violation of Hara's
theorem results from breaking either gauge- or CP- invariance
in quark-level calculations.  Quark-model calculations are, however,
explicitly gauge- and CP- invariant, whether one uses the potential
model \cite{ref:KR} or the bag model \cite{ref:Lo}.
The emerging question is thus: How can the quark model satisfy gauge-
and CP- invariance, and yet violate the theorem?

\section{AWAY FROM $SU(3)$}
In the past an additional problem was caused by the sign of the
\Sidec\ asymmetry.  Namely, assuming that the \Sidec\ decay is
dominated by the single-quark diagram of Fig. 2a, one can show
\cite{ref:Lach,ref:Vasanti}
that asymmetry in question is
\begin{equation}
\alpha (\Sigma ^+ \rightarrow p \gamma ) = \frac{m^2_s - m^2_d}{m^2_s +
m^2_d}
\end{equation}
which is positive (+0.4 or +1.0 for
constituent or current quark masses respectively) and thus in
disagreement with experiment.
Recent precise measurements of the $\Xi ^- \rightarrow \Sigma ^- \gamma $
branching ratio \cite{ref:BRXi}
(which proceeds through diagram $(2a)$ only)
prove, however, that there is no way of reproducing
the \Sidec\ branching ratio by assuming the dominance
of diagram $(2a)$: the predicted branching ratio is then
too small by a factor of one hundred.

\section{QUARK DIAGRAMS}
Out of all topologically possible quark diagrams shown in Fig. 2,
contribution
from diagrams $(c)$ vanishes in the $SU(3)$ limit and is negligible
in explicit calculations with broken $SU(3)$
\cite{ref:Lach,ref:Gavela}.
Diagrams $(d)$ are suppressed by the presence of two $W$ propagators.
Thus, it is contribution from diagrams $(b1)$ and $(b2)$
{\em only} that may be significant.
Violation of Hara's theorem results from this very set
of quark diagrams \cite{ref:KR}.
\subsection{Hadron-level way}
At the hadron level diagrams $(b1)$ and $(b2)$ correspond to the
contribution from intermediate $\frac{1}{2}^-$ excited baryons.
Using the quark model one can calculate the $\frac{1}{2}^+
\rightarrow
\frac{1}{2}^-$ weak transition elements and the
$\frac{1}{2}^-\rightarrow \frac{1}{2}^+ + \gamma $
electromagnetic couplings.  Their relative size is governed by
group-theoretical
spin-flavour factors, the products of which are given in Table 1.
When one \underline{identifies} the results of these quark model
calculations with {\em those hadron-level} expressions that are
{\em allowed by gauge invariance},
one finds that contributions from diagrams $(b1)$  and $(b2)$ must
enter with a relative minus sign, thus ensuring cancellation
(in the $SU(3)$ limit) of the corresponding
contributions to the \Sidec\ decay\cite{ref:Gavela}.
\\
\\
\setlength{\unitlength}{0.43pt}
\begin{picture}(730,520)
\put(230,0){\begin{picture}(430,520)

\put(115,360){
\begin{picture}(200,160)
\put(100,15){\makebox(0,0){(a)}}
\put(30,90){\line(1,0){25}}
\put(100,90){\vector(-1,0){45}}
\put(100,90){\line(1,0){45}}
\put(170,90){\vector(-1,0){25}}
\put(185,93){s}
\put(5,93){d}
\put(20,130){$\gamma $}

\multiput(95,90)(-10,10){5}{\put(0,0) {\oval(10,10)[tr]}
                             \put(0,10){\oval(10,10)[bl]}}
\put(100,90){\circle*{10}}

\put(170,65){\vector(-1,0){70}}
\put(100,65){\line(-1,0){70}}
\put(170,40){\vector(-1,0){70}}
\put(100,40){\line(-1,0){70}}
\multiput(85,90)(5,0){6}{\line(1,0){3}}
\end{picture}}
\put(0,0){
\begin{picture}(200,160)
\put(100,15){\makebox(0,0){(c)}}
\put(30,90){\line(1,0){25}}
\put(100,90){\vector(-1,0){45}}
\put(100,90){\line(1,0){45}}
\put(170,90){\vector(-1,0){25}}

\multiput(95,90)(-10,10){5}{\put(0,0) {\oval(10,10)[tr]}
                             \put(0,10){\oval(10,10)[bl]}}
\put(20,130){$\gamma $}

\put(170,65){\vector(-1,0){70}}
\put(100,65){\line(-1,0){70}}
\put(170,40){\vector(-1,0){70}}
\put(100,40){\line(-1,0){70}}
\multiput(100,40)(0,5){5}{\line(0,1){3}}
\put(105,45){$W$}
\end{picture}}
\put(230,0){
\begin{picture}(200,160)
\put(100,15){\makebox(0,0){(d)}}
\put(30,90){\line(1,0){25}}
\put(100,90){\vector(-1,0){45}}
\put(100,90){\line(1,0){45}}
\put(170,90){\vector(-1,0){25}}

\put(170,65){\vector(-1,0){70}}
\put(100,65){\line(-1,0){70}}
\put(170,40){\vector(-1,0){70}}
\put(100,40){\line(-1,0){70}}
\multiput(100,65)(0,5){5}{\line(0,1){3}}
\put(105,70){$W$}
\multiput(100,77.5)(-10,10){5}{\put(0,5) {\oval(10,10)[bl]}
                               \put(-10,5){\oval(10,10)[tr]}}
\put(20,117.5){$\gamma $}
\end{picture}}

\put(0,180){
\begin{picture}(200,160)
\put(100,15){\makebox(0,0){(b1)}}
\put(170,90){\vector(-1,0){25}}
\put(145,90){\line(-1,0){65}}
\multiput(130,65)(0,5){5}{\line(0,1){3}}
\put(135,70){$W$}
\put(80,90){\vector(-1,0){35}}
\put(45,90){\line(-1,0){15}}

\multiput(80,90)(-10,10){5}{\put(0,0) {\oval(10,10)[tr]}
                             \put(0,10){\oval(10,10)[bl]}}
\put(0,130){$\gamma $}

\put(170,65){\vector(-1,0){90}}
\put(80,65){\line(-1,0){50}}
\put(170,40){\vector(-1,0){90}}
\put(80,40){\line(-1,0){50}}
\end{picture}}
\put(230,180){
\begin{picture}(200,160)
\put(100,15){\makebox(0,0){(b2)}}
\put(170,90){\vector(-1,0){15}}
\put(155,90){\line(-1,0){35}}

\multiput(120,90)(-10,10){5}{\put(0,0) {\oval(10,10)[tr]}
                             \put(0,10){\oval(10,10)[bl]}}
\put(40,130){$\gamma $}

\put(170,65){\vector(-1,0){50}}
\put(120,65){\line(-1,0){90}}
\put(170,40){\vector(-1,0){50}}
\put(120,40){\line(-1,0){90}}
\put(120,90){\vector(-1,0){65}}
\put(55,90){\line(-1,0){25}}
\multiput(70,65)(0,5){5}{\line(0,1){3}}
\put(75,70){$W$}
\end{picture}}
\end{picture}
}\end{picture}
\\
\\
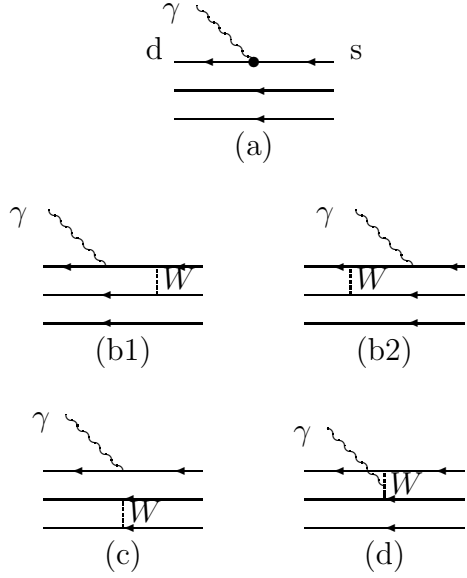
\begin{center}Fig. 2. Quark diagrams for weak radiative
hyperon decays.
\end{center}

In explicit models $SU(3)$ is broken in energy denominators
resulting from propagation of the intermediate excited
$\frac{1}{2}^-$ baryon.
Since $m_{N^*} - m_{\Sigma ^+} = \Delta \omega - \delta s$,
$m_{\Sigma ^*} - m _p = \Delta \omega + \delta s $ (where
$\Delta \omega \approx 0.57 GeV$ is the energy difference between
excited and ground-state baryons, and
$\delta s = m_s
- m_{u,d} \approx 0.19 GeV $ is the strange-nonstrange quark mass
difference), diagrams $(b1)$ and $(b2)$ - having different energy
denominators - do not cancel exactly \cite{ref:Gavela}.
The corresponding formulae (up to an uninteresting normalization
factor) are given in column 2 of Table 2, where $x \equiv
\frac{\delta s}{\Delta \omega } \approx \frac{1}{3}$. By construction
the obtained \Sidec\ parity violating amplitude vanishes in
the $SU(3)$ limit ($x \rightarrow 0$).
\\
\\
\\
Table 1. Group-theoretical factors for diagrams $(b1)$ and $(b2)$
\begin{footnotesize}
\begin{center}
\begin{tabular}{l c c}
\hline
&&\\
process & diag. $(b1)$ & diag. $(b2)$       \\
\hline
&&\\
$\Sigma ^+ \rightarrow p \gamma $ &
\begin{normalsize} $-\frac{1}{3\sqrt{2}}$ \end{normalsize} &
\begin{normalsize} $-\frac{1}{3\sqrt{2}}$ \end{normalsize} \\
&&\\
$\Lambda \rightarrow n \gamma $ &
\begin{normalsize} $+\frac{1}{6\sqrt{3}}$ \end{normalsize} &
\begin{normalsize} $+\frac{1}{2\sqrt{3}}$ \end{normalsize} \\
&&\\
$\Xi ^0 \rightarrow \Lambda \gamma $ &
\begin{normalsize} $0$ \end{normalsize}&
\begin{normalsize} $-\frac{1}{3\sqrt{3}}$ \end{normalsize} \\
&&\\
$\Xi ^0 \rightarrow \Sigma ^0 \gamma $ &
\begin{normalsize} $+\frac{1}{3}$ \end{normalsize} &
\begin{normalsize} $0$ \end{normalsize} \\
&&\\
\hline
\end{tabular}
\end{center}
\end{footnotesize}

\subsection{Quark-level way}
There is, however, no reason to identify quark model calculations
of the $\frac{1}{2}^+ \rightarrow \frac{1}{2}^-$
weak transition elements and the $\frac{1}{2}^- \rightarrow
\frac{1}{2}^+ + \gamma $ electromagnetic couplings with
hadron-level expressions.  One can perform all calculations at the
strict quark level and only eventually evaluate the resulting
expression in between the initial and final hadronic states.
These direct quark-model calculations (potential model \cite{ref:KR},
bag model \cite{ref:Lo}) yield amplitudes proportional to the
\underline{sum} of spin-flavour factors corresponding to diagrams
$(b1)$ and $(b2)$.
In a consistent quark-level calculation the relative sign of
spin-flavour factors of diagrams $(b1)$ and $(b2)$ is obviously
{\em fixed} and it turns out to be {\em positive}: Energy
denominators corresponding to diagrams $(b1)$ and $(b2)$ are of the
{\em same} sign. With a relative positive sign the
contributions of diagrams $(b1)$ and $(b2)$ add rather than cancel
resulting in the violation of Hara's theorem (see column 3, Table
2). Therefore, it is through \underline{insistence} on identifying
quark-level expressions with the {\em hadron-level
gauge-invariance-allowed} amplitudes \underline{only} that the relative
negative sign was previously generated.
\\
\\
\\
Table 2. Parity violating amplitudes with SU(3) breaking:
\\
\phantom{xxxxx}$(b1)$ - $(b2)$ - Hara's theorem satisfied,
\\
\phantom{xxxxx}$(b1)$+$(b2)$ - Hara's theorem violated.
\begin{footnotesize}
\begin{center}
\begin{tabular}{lcc}
\hline
&&\\
process & \phantom{x}$(b1)$-$(b2)$ & \phantom{x}$(b1)$+$(b2)$ \\
\hline
&&\\
$\Sigma ^+ \rightarrow p \gamma $ &
\begin{normalsize} $-\frac{2x}{3\sqrt{2}}$ \end{normalsize} &
\begin{normalsize} $-\frac{2}{3\sqrt{2}}$ \end{normalsize} \\
&&\\
$\Lambda \rightarrow n \gamma $ &
\begin{normalsize} $+\frac{2x-1}{3\sqrt{3}}$ \end{normalsize} &
\begin{normalsize} $+\frac{2-x}{3\sqrt{3}}$ \end{normalsize} \\
&&\\
$\Xi ^0 \rightarrow \Lambda \gamma $ &
\begin{normalsize} $+\frac{1-x}{3\sqrt{3}}$ \end{normalsize} &
\begin{normalsize} $-\frac{1-x}{3\sqrt{3}}$ \end{normalsize} \\
&&\\
$\Xi ^0 \rightarrow \Sigma ^0 \gamma $ &
\begin{normalsize} $+\frac{1+x}{3}$ \end{normalsize} &
\begin{normalsize} $+\frac{1+x}{3}$ \end{normalsize} \\
&&\\
\hline
\end{tabular}
\end{center}
\end{footnotesize}

\section{A CLOSER LOOK}
The problem is thus as follows:
\begin{itemize}
\item if we apply gauge invariance at \underline{hadron level}
(original proof of Hara's theorem, or pole model
with $\frac{1}{2}^-$ intermediate baryons)
- Hara's theorem is \underline{satisfied}
\item if we apply gauge invariance at \underline{quark level} - Hara's
theorem is \underline{violated}.
\end{itemize}

In order to understand this one needs a way to {\em translate}
the gauge-invariance condition from the quark to the hadron level (instead
of using an {\em ad hoc} identification prescription). The way to
do it is called the Kroll-Lee-Zumino
scheme \cite{ref:KLZ}.  According to the KLZ scheme, translation
of quark-level interactions with a photon to the hadron-level
language is provided by the vector dominance model (VDM).

Standard VDM prescription is formulated at the {\em hadron
level} and consists in:
\begin{enumerate}
\item calculating vector meson ($V^{\mu }$) couplings to hadrons ($H_1,H_2$)
through
$<~H_2|J^V_{\mu }|H_1~> V^{\mu }$
where $J^{V}_{\mu }$ are quark currents
\item replacing vector mesons by photons through
$V^{\mu } \rightarrow \frac{e}{g_V} A^{\mu }$ (where $g_{\rho } = 5.0$).
\end{enumerate}
The latter step may be obtained at a theoretical level by
introducing
a gauge-invariance-violating coupling $e \frac{m^2_V}{g_V} V\cdot A$
that induces photon mass.  In the KLZ scheme one adds additional
terms to cancel this photon mass so that gauge invariance is restored.
Then, after redefining
photon and vector-meson fields as well as
electric charge,
the VDM prescription turns out to be just a {\em good approximation
to the quark-level} prescription in which photons
couple to quarks directly and in an explicitly gauge-invariant way:
$<H_2|J^V_{\mu }|H_1> A^{\mu }$
(for details see \cite{ref:Lach,ref:KLZ,ref:Sakurai}).

The KLZ scheme permits an understanding of the origin of
the violation of Hara's theorem in the quark model \cite{ref:Zen89}.
Namely, explicit calculations of diagrams $(b1)$  and $(b2)$ with
photon
replaced by vector meson show that the coupling $\Sigma ^+ \rightarrow p$
$+$
$(U$-$spin$ $singlet$ $vector$ $meson)$ does \underline{not} vanish.
Since no gauge-invariance condition is imposed in vector-meson case it
is clear that the obtained coupling may be identified with the
$F_1(q^2) \overline{\Psi }_p \gamma _5 \gamma _{\mu }\Psi _{\Sigma ^+}
V^{\mu }$ term with a nonvanishing $F_1(0)$.  This is in fact the
standard identification (see eg. references contained in ref.
\cite{ref:Lach}).
Thus, according to the KLZ scheme and ref.\cite{ref:Zen89} the
quark-model result corresponds to the VDM-generated
effective coupling $F_1(0) \overline{\Psi }_p \gamma _5
\gamma _{\mu }\Psi _{\Sigma ^+} A^{\mu }$ that does \underline{not}
vanish at
$q^2 = 0$.
This coupling was absent in the original derivation of Hara's theorem
in which, therefore,
{\em contribution
from pointlike quarks was simply not taken into account}.

\section{OBSERVABLE CONSEQUENCES}
When parity-violating amplitudes of Table 2 are supplemented with
standard description of parity-conserving amplitudes one obtains
different signatures for hadron- and quark- level predictions
(see Table 3).
Namely, if Hara's theorem is satisfied (as in hadron-level approaches)
{\em all} four asymmetries are of the {\em same} sign.
On the other hand, if the quark-model route is strictly followed,
Hara's theorem is violated and the asymmetries of the
$\Lambda \rightarrow n \gamma $ and the $\Xi ^0 \rightarrow \Lambda
\gamma $ decays are
{\em opposite} to those of \Sidec\ and $\Xi ^0 \rightarrow \Sigma ^0
\gamma $.
Phenomenologically, the $\Xi ^0 \rightarrow \Lambda \gamma $ decay is a
much cleaner case than $\Lambda \rightarrow n \gamma $ (see
ref.\cite{ref:Lach}).
It is therefore extremely important that the asymmetry of the
$\Xi ^0 \rightarrow \Lambda \gamma $ be precisely measured.  Current
data
(Table 3) on the $\Xi ^0 \rightarrow \Lambda \gamma $ asymmetry
reject
Hara's theorem
at an almost $3 \sigma $ level.
When other asymmetries and branching ratios are taken into account
the disagreement with Hara's theorem is even more significant
(Table 3, for full account see ref.\cite{ref:Lach}).

We are therefore eagerly awaiting the results of the hyperon decay
program in the E832 KTeV experiment at Fermilab, where the expected
number of $\Xi ^0 \rightarrow \Lambda \gamma $ events is 900, a factor
of 10 greater than the number of events observed thus far.  Measurements
of the $\Xi ^0 \rightarrow \Sigma ^0 \gamma $ asymmetry, planned in
the same experiment,
are also important: for this decay {\em all} models predict
negative
(and often large) asymmetries while the only experiment performed so far
does not support a large negative asymmetry.
\\
\\
\\
Table 3. Asymmetries and branching ratios - comparison of two
selected conflicting models and experiment
\begin{footnotesize}
\begin{center}
\begin{tabular}{lccc}
\multicolumn{4}{c}{Asymmetries}
\\
\hline
process & ref. \cite{ref:Gavela} & exp. & ref.
\cite{ref:Lach} \\
& Hara th. & & Hara th. \\
& satisfied & & violated \\
\hline
&&&\\
$\Sigma ^+ \rightarrow p \gamma $ & $-0.80 ^{+0.32}_{-0.19}$&
$-0.76 \pm 0.08$ & $-0.95$ \\
$\Lambda \rightarrow n \gamma $ & $-0.49$ &
& $+0.80$ \\
$\Xi ^0 \rightarrow \Lambda \gamma $ & $-0.78$ &
$+0.43 \pm 0.44$ & $+0.80$ \\
$\Xi ^0 \rightarrow \Sigma ^0 \gamma $ & $-0.96$ &
$+0.20 \pm 0.32$ & $-0.45$ \\
\hline
&&&\\
\multicolumn{4}{c}{Branching ratios (in units of $10^{-3}$)}
\\
\hline
&&&\\
$\Sigma ^+ \rightarrow p \gamma $ & $0.92 ^{+0.26}_{-0.14}$&
$\phantom{+}1.23 \pm 0.06$ & $1.3-1.4$ \\
$\Lambda \rightarrow n \gamma $ & $0.62$ &
$\phantom{+}1.63 \pm 0.14$ & $1.4-1.7$ \\
$\Xi ^0 \rightarrow \Lambda \gamma $ & $3.0$ &
$\phantom{+}1.06 \pm 0.16$ & $0.9-1.0$ \\
$\Xi ^0 \rightarrow \Sigma ^0 \gamma $ & $7.2$ &
$\phantom{+}3.56 \pm 0.43$ & $4.0-4.1$ \\
\hline
\end{tabular}
\end{center}
\end{footnotesize}
\section{LOOKING DEEPER}
I believe that in a few years' time predictions of the quark and
vector-dominance models
will be better confirmed experimentally.  The problem will then be
to understand this result at a deeper theoretical level.

Technical reasons for the difference between the
original hadron-level predictions and the quark or vector-dominance
models are already obvious.
Namely, in the most naive quark-level calculations quarks are
treated as free particles subject to proper
(anti)symmetrization of their total wave function.
Clearly, the gauge-invariance condition imposed in this language
(with gauge transformations on quark fields located at $x_1,x_2,x_3$)
is not equivalent to the gauge-invariance condition imposed in
the hadron-level language (where gauge transformations are performed
on an effective hadron field located at a different point
$x$).
When such free quarks are confined by phenomenological tools
(as in eg. potential model, bag model etc.) the difference in question
does \underline{not} vanish. In particular,
unless artificially tailored to satisfy the standard hadron-level
gauge-invariance condition, all QCD-inspired approaches with built-in
contribution of {\em free} quarks must also yield violation of Hara's
theorem.

The physical origin of problems with Hara's theorem is therefore
related
to the issue of unobservability of apparently free quarks.
Violation of Hara's theorem by
the quark and vector-dominance models indicates that our present
understanding of this point is very unsatisfactory.
This question has been with us since the beginnings of the quark model
(cf. the dubious assumption of additivity of magnetic moments of
Dirac quarks which are free and yet always grouped into hadrons).
Since the quark model was so tremendously successful,
ways of maintaining the contribution from free quarks have been
proposed that keep in line with the apparent unobservability of quarks
in asymptotic states. With the advent of precise measurements
of WRHD's the original questions reappear with greatly increased
strength. I do not think one can answer them in the traditional way:
these questions appear in any QCD-inspired quark-confining framework
with built-in contribution from free quarks.
Ways of representing the freedom of quarks, different from the
current ones, would have to be devised should Hara's theorem be
satisfied
and quark freedom maintained.  Such attempts would then have to confront
the ultimate judge - the experiment.  The latter favours
the violation of Hara's theorem, though.

Problems with Hara's theorem are clearly related to a space-time
description of composite hadronic states.
It is therefore very interesting to note that
the case of composite quantum states is
beset with conceptual problems at the quantum/special relativity
interface, problems that appear at any distance scale.
In the opinion of many physicists working on the foundations of physics
these
problems require a profound change in our understanding of the nature of
space.
Thus it is very intriguing to note that the KLZ scheme
may be viewed as connecting
alternative
"space representations" of the underlying physics:
the descriptions in terms of constituents
(quarks located at
points $x_1,x_2,x_3$) and those in terms of composites
(hadron located at $x$).
In my opinion, therefore, hadron physics is more intimately
related to the nature of space than it is generally acknowledged.

\section{SUMMARY}
In summary,  WRHD's probe the {\em very
basic} {\em assumptions} of the quark model. These
assumptions are {\em in direct conflict} with
the standard way of imposing gauge-invariance
condition at hadron level.

{\em One cannot have both.}
One must either drastically modify the basic assumptions of the
quark model or admit that the standard way of imposing the
gauge-invariance condition at hadron level does not have much to do
with what happens in Nature.
Is the quark-model way correct indeed?
And - if yes - what does it mean?

I believe that WRHD's provide us with an important clue to
a deeper understanding of the
question of how apparently free quarks combine to form
hadrons as the only observable asymptotic states.
\\

This work is supported in part by the KBN grant No 2P0B23108.

\end{document}